\UseRawInputEncoding
\documentclass[reprint,aps,prb,floatfix,twocolumn,superscriptaddress,showpacs,amsmath,amssymb,showpacs,]{revtex4-1} 

\usepackage{graphicx}
\usepackage{epstopdf}
\usepackage{epsfig}
\usepackage{epsf}
\usepackage{url}
\usepackage[USenglish]{babel}
\usepackage{hyperref}
\def\bcen{\begin{center}}
\def\ecen{\end{center}}
\allowdisplaybreaks
\renewcommand\[{\begin{equation}}
\renewcommand\]{\end{equation}}
\usepackage{verbatim}
\usepackage{xcolor}
\usepackage{amsmath, nccmath}
\usepackage{bm}
\usepackage{bbm}
\begin{document}
\title{Fully {\it ab-initio} electronic structure of Ca$_{2}$RuO$_{4}$}
\author{Francesco Petocchi}
\affiliation{Department of Physics, University of Fribourg, 1700 Fribourg, Switzerland}
\author{Viktor Christiansson}
\affiliation{Department of Physics, University of Fribourg, 1700 Fribourg, Switzerland}
\author{Philipp Werner}
\affiliation{Department of Physics, University of Fribourg, 1700 Fribourg, Switzerland}
\begin{abstract}
The reliable {\it ab-initio} description of strongly correlated materials is a long-sought capability in condensed matter physics. The $GW$+EDMFT method is a promising scheme, which provides a self-consistent description of correlations and screening, and does not require user-provided parameters. In order to test the reliability of this approach we apply it to the experimentally well characterized perovskite compound Ca$_2$RuO$_4$, in which a temperature-dependent structural deformation drives a paramagnetic metal-insulator transition. Our results demonstrate that the nonlocal polarization and self-energy components introduced by $GW$ are essential for setting the correct balance between interactions and bandwidths, and that the $GW$+EDMFT scheme produces remarkably accurate predictions of the electronic properties of this strongly correlated material.
\end{abstract}
\date{\today}
\maketitle
%
%
%
%
\section{INTRODUCTION}
Density functional theory (DFT)\cite{Hohenberg1964,Kohn1965} does not provide an adequate description of the electronic structure of strongly correlated materials, but substantial progress in the simulation of this class of materials has been made by adding the missing correlations via some dynamical mean-field theory (DMFT) construction.\cite{Metzner1989,Georges1992,Georges1996} The widely used DFT+DMFT\cite{Kotliar2006} scheme however has several drawbacks which prevent {\it ab-initio} predictions. In particular, a true {\it ab-initio} procedure needs to provide the interaction parameters that go into the DMFT calculation, and the latter should be consistent with the correlated electronic structure. A more advanced scheme is the combination of extended (E)DMFT\cite{Sun2002} with the $GW$\cite{Hedin1965} method ($GW$+EDMFT),\cite{Biermann2003,Ayral2013,Werner2016} which provides a self-consistent description of correlations and screening in a solid. Apart from the choice of a subspace of strongly correlated orbitals, $GW$+EDMFT is free from ad-hoc parameters and thus represents a true {\it ab-initio} method for strongly correlated materials. 

While first applications of this recently developed simulation framework have produced promising results,\cite{Boehnke2016,Nilsson2017,Petocchi2019,Petocchi2020} the reliability of $GW$+EDMFT still needs to be systematically tested. A particularly relevant question is if the self-consistently computed effective local interaction parameters, which are obtained by several downfolding steps,\cite{Nilsson2017} correctly capture the correlation strength in the material. To clarify this point, it is useful to investigate a compound in which a correlation-driven metal-insulator transition (MIT) occurs without a simultaneous electronic ordering transition, since this allows to disentangle the opening of a Mott gap from, e.~g., the effect of antiferromagnetic correlations. Here, we study the layered perovskite Ca$_2$RuO$_4$, which exhibits a transition between a metal and Mott insulator (and a simultaneous structural transition) at a temperature near 360~K,\cite{Braden1998,Alexander1999} which is significantly larger than the N\'eel temperature of the antiferromagnetic (AFM) phase emerging below $T_N\sim$110~K.\cite{Braden1998} As we will show, $GW$+EDMFT predicts solutions for the two structures which are in good agreement with the available photo-emission data,\cite{Sutter2017,Miyashita2021} and no metal-insulator transition near the experimental temperature in the absence of a structural transition. In contrast, an EDMFT treatment based on the downfolded interactions would predict insulating solutions with a far too large gap for both structures. This demonstrates the important role of nonlocal screening processes in producing physically correct interaction strengths within self-consistent $GW$+EDMFT.  

The paper is organized as follows: In Sec.~\ref{section:methods} we summarize the $GW$+EDMFT method and explain its extension to materials with multiple correlated sites within the unit cell. In Sec.~\ref{section:Results} we present the results obtained for the two studied crystal structures and provide comparisons with available photoemission data. In Sec.~\ref{section:Conclusions} we present our conclusions.
%
%
%
\section{METHOD}\label{section:methods}
\subsection{DFT and single-shot $G_{0}W_{0}$}\label{subsection:DFT}
\begin{figure}
\begin{center}
\hspace{0.55cm}
\includegraphics[width=0.4\textwidth]{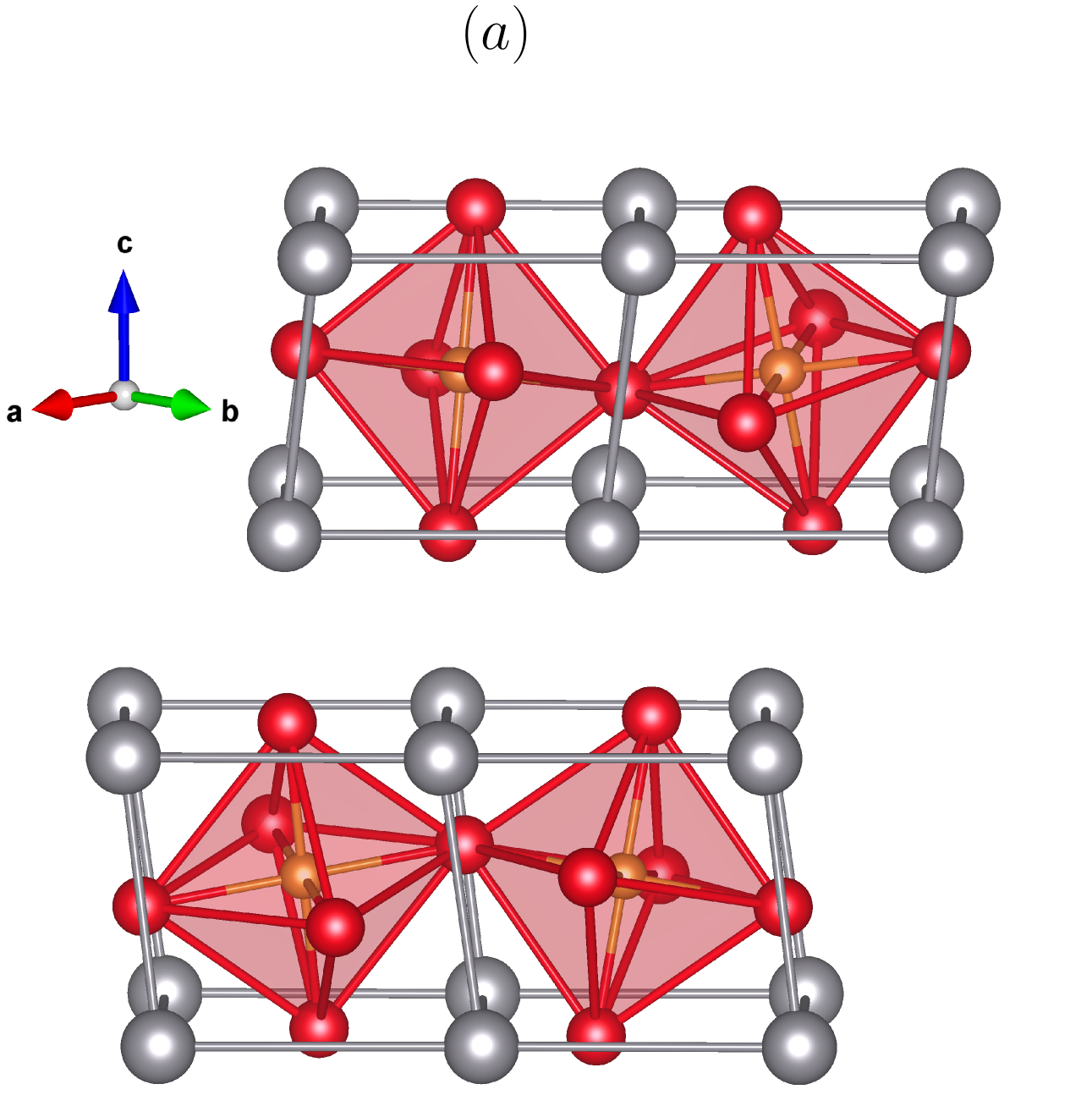}
\vspace{0.4cm}\\
\includegraphics[width=0.28\textwidth]{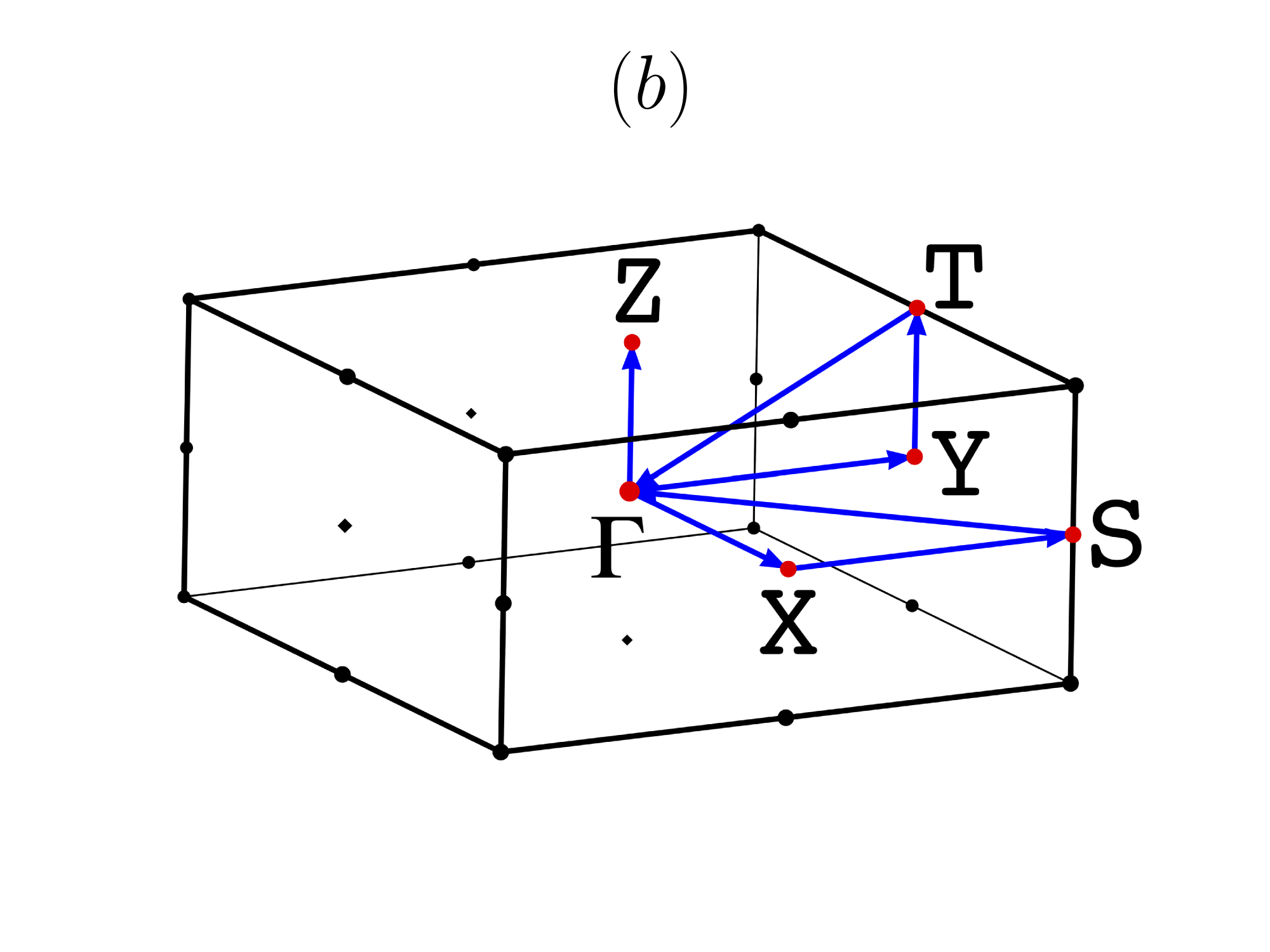}
\caption{(a) \textit{S}-Pbca crystal structure drawn using the VESTA software.\cite{Momma2011} The primitive unit cell consists of four formula units. The Ru atoms (orange) are inside tilted octahedra of O (red) surrounded by a cage of Ca (gray).
(b) First Brillouin zone and high-symmetry lines.\label{Figure: Structure}}
\end{center}
\end{figure}

The DFT calculations were done using the full-potential linearized augmented plane-wave code FLEUR \cite{Fleurcode} and a 16$\times$16$\times$16 ${\mathbf{k}}$-grid. The space group of Ca$_2$RuO$_4$ is Pbca, and the lattice parameters as well as the positions of the atoms are taken to be the experimental ones from Ref.~\onlinecite{Friedt2001}. The lattice parameters are given in Table \ref{Table: Wyckoff Position} for the long ($L$) and short ($S$) cell, with the latter structure shown in Fig.~\ref{Figure: Structure} together with the corresponding first Brillouin zone. The crystal structure consists of Ru atoms inside tilted and rotated O octahedra, surrounded by a cage of Ca atoms. The long cell is characterized by a longer $c$-axis and a slightly increased rotation angle, while the tilt angle is reduced compared to the short cell.\cite{Friedt2001} The Wyckoff positions are for Ru: (4a)-(0,0,0), and for Ca, O(1), O(2): (8c)-(x,y,z) with the positional parameters given in Table \ref{Table: Wyckoff Position}. Here, O(1) and O(2) refers to the two inequivalent oxygen positions.
\begin{table}[ht!]
\setlength{\tabcolsep}{15pt} 
\renewcommand{\arraystretch}{1.5} 
\centering
\begin{tabular}{|c|c|c|c|}
\hline
\multicolumn{1}{c}{} & \multicolumn{1}{c}{x}  & \multicolumn{1}{c}{y}  & \multicolumn{1}{c}{z} \\
\hline
\multicolumn{4}{c}{\textit{S}-Pbca (a=5.3945, b=5.5999, c=11.7653)}    \\
\hline
Ca & 0.0042 & 0.0559 & 0.3524 \\
O(1) & 0.1961 & 0.3018 & 0.0264 \\
O(2) & -􏰔0.0673 & -0.0218 &0.1645 \\
\hline
\multicolumn{4}{c}{\textit{L}-Pbca: (a=5.3606, b=5.3507, c=12.2637)}  \\
\hline
Ca & 0.0110 & 0.0269 & 0.3479 \\
O(1) & 0.1939 & 0.3064 & 0.0147 \\
O(2) & -0.0386 & -0.0067 & 0.1656 \\
\hline
\end{tabular}
\caption{Lattice parameters in \AA \, and position parameters for the Wyckoff positions for the long ($L$) and short ($S$) cell of Ca$_2$RuO$_4$. O(1) and O(2) represent the two inequivalent oxygen positions.\label{Table: Wyckoff Position}}
\end{table}
We use the Wannier90 library\cite{Mostofi2008} to define a low-energy model for the bands around the Fermi energy with three maximally localized Wannier functions\cite{Marzari1997} centered on each of the four Ru sites. For the constrained random phase approximation (cRPA)\cite{Aryasetiawan2004} and single-shot $G_0W_0$ calculations we use the SPEX code,\cite{Friedrich2010} where we employ a 7$\times$7$\times$5 ${\mathbf{k}}$-grid together with 300 bands in the calculation of both the polarization and self-energy.
\subsection{Real space extension of $GW$+EDMFT}\label{subsection:Real_space_GW+EDMFT}
The {\it ab-initio} $GW$+EDMFT method for correlated materials has been developed and described in Refs.~\onlinecite{Boehnke2016,Nilsson2017,Petocchi2019,Petocchi2020}. In contrast to the widely-used DFT+DMFT approach,\cite{Kotliar2006} it is fully {\it ab-initio} and does not require ad-hoc parameters such as local interactions and double-counting corrections. The only choice which needs to be made is the definition of physically motivated \mbox{high-,} intermediate-, and low-energy subspaces (or ``tiers" \cite{Nilsson2017}), in which correlations will be treated at the $G_0W_0$, self-consistent $GW$, and EDMFT level, respectively. In the present case of Ca$_2$RuO$_4$ we treat in the intermediate and correlated low-energy subspace the $4d$ $t_{2g}$ orbitals of the Ru atoms, and treat the higher-energy electronic degrees of freedom at the $G_0W_0$ level. 

Spin-orbit coupling (SOC) is known to be relevant for Ruthenates,\cite{Mackenzie1996} and the resulting hybridized $t_{2g}$ orbitals are usually referred to as $\alpha$, $\beta$ and $\gamma$ multiplets. In the present study, which is restricted to the high-temperature phases of Ca$_2$RuO$_4$, we however neglect SOC, so that the orbitals retain their $t_{2g}$ character. 

The $GW$+EDMFT simulation starts with a DFT calculation in the local density approximation (LDA),\cite{Kohn1965} as described in Sec.~\ref{subsection:DFT}. By means of a single-shot $G_0W_0$ calculation, the effective ``bare" interactions $U_{\bf q}^\text{cRPA}$ and the effective ``bare" lattice Greens functions $G_{\bf k}^{G_0W_0}$ of the intermediate/low-energy space are computed. With these, the self-consistent $GW$+EDMFT calculation is performed in the intermediate/low-energy space, starting from a vanishing EDMFT polarization and self-energy ($\Pi^{\mathrm{EDMFT}}=0$, $\Sigma^{\mathrm{EDMFT}}=0$). The self-consistency loop fixes the EDMFT impurity problem with fermionic and bosonic baths such that the local lattice Green's function is identical to the impurity Green's function, $G_{\mathrm{loc}}=G_{\mathrm{imp}}$ (fermionic self-consistency condition), and that the screened local lattice interaction is identical to the screened impurity interaction, $W_{\mathrm{loc}}=W_{\mathrm{imp}}$ (bosonic self-consistency condition).\cite{Sun2002} This is imposed for each site within the unit cell that is treated at the EDMFT level, and hence mapped onto an impurity problem with retarded effective interaction (here, the four Ru atoms).  The local EDMFT self-energies and polarizations replace the corresponding local $GW$ contributions, and thereby strong correlation effects can be incorporated into the simulation without ill-defined double countings. The method is able to treat different types of correlated atoms in a unit cell, and to couple them at the $GW$ level, as demonstrated in Ref.~\onlinecite{Petocchi2020}. 

In the case of Ca$_2$RuO$_4$, we have to consider correlated orbitals residing on ions of the same species (Ru), but inequivalently arranged in the unit cell (Fig.~\ref{Figure: Structure}). In this situation, the local Hamiltonians of the different sites $\mathcal{H}_{\mathrm{loc}}^{i}$ become equivalent provided that the local problem is rotated to the diagonal or crystal-field basis via a site-dependent rotation $\Theta^{i}$. In the crystal-field basis the local noninteracting Hamiltonians are replaced by local energies $\varepsilon_{\mathrm{loc}}^{i}$, which are identical for each correlated site, and hence only a single impurity problem needs to be solved. Before the self-consistent calculation is started, we determine the rotations $\Theta^{i}$ 
\begin{equation}
\hat{\Theta}=\left(\begin{array}{c}
\left[\Theta^{1}\right]\\
\ldots\\
\left[\Theta^{N_{s}}\right]
\end{array}\right),
\end{equation}
which provide the mapping between the original and the crystal-field basis. 
In the present case, $N_{s}=4$ is the number of Ru sites in the unit cell. If the interaction is of the Kanamori form and restricted to the $t_{2g}$ or $e_{g}$ manifold, no transformation is needed in the case of a rotationally invariant formulation ($J=J_{X}=J_{P}$, see Ref.~\onlinecite{Georges2013}). This follows from the fact that a $3\times3$ real rotation is preserving the SO$\left(3\right)$ invariance of the interaction.\cite{Georges2013} For the full cRPA tensor that is used as a bare interaction in $GW$+EDMFT, this is generally not the case, especially when the orbitals are slightly splitted already at the LDA level as in Ca$_{2}$RuO$_{4}$. In such systems, one needs to apply the tensor analogue of the fermionic matrix rotation to the local effective bosonic interactions related to each site. A brief summary of this procedure is provided in the Appendix. As it is the case for the local energies and hybridization functions, in the crystal field basis, the transformed interaction tensors become equivalent among the sites, thus ensuring the invariance of the impurity models. 

In the following, we summarize the steps in the $GW$+EDMFT self-consistency loop for locally equivalent sites:
\begin{enumerate}
\item Compute the $\mathbf{k}$ dependent self-energy and polarization in the $GW$ approximation (polarization $\Pi_{\mathbf{q}}^{GG}$ and self-energy $\Sigma_{\mathbf{k}}^{GW}$, see Ref.~\onlinecite{Nilsson2017} for details) and replace the local components with the EDMFT results:
\begin{itemize}
\item $\Pi_{\mathbf{q}}=\Pi_{\mathbf{q}}^{GG}-\Pi_{\mathbf{q}}^{GG}|_{\mathrm{loc}}+\Pi^{\mathrm{EDMFT}}$,
\item $\Sigma_{\mathbf{k}}=\Sigma_{\mathbf{k}}^{GW}-\Sigma_{\mathbf{k}}^{GW}|_{\mathrm{loc}}+\Sigma^{\mathrm{EDMFT}}$.
\end{itemize}
\item Use the polarization and self-energy, as well as the bare propagators $U_{\mathbf{q}}^\text{cRPA}$ and $G_{\mathbf{k}}^{G_0W_0}$, to extract the local screened interaction and local lattice Green's function:
\begin{itemize}
\item $W_{\mathrm{loc}}=\sum_{\mathbf{q}}U_{\mathbf{q}}^\text{cRPA}\left[1-\Pi_{\mathbf{q}}U_{\mathbf{q}}^\text{cRPA}\right]^{-1}$,
\item $G_{\mathrm{loc}}=\sum_{\mathbf{k}}\left[\left(G_{\mathbf{k}}^{G_0W_0}\right)^{-1}-\Sigma_{\mathbf{k}}\right]^{-1}$.
\end{itemize}
\item Impose the two self-consistency conditions (at some given site $i$):
\begin{itemize}
\item $W_{\mathrm{loc}}^{i}=W_{\mathrm{imp}}^{i}$,
\item $G_{\mathrm{loc}}^{i}=G_{\mathrm{imp}}^{i}$.
\end{itemize}
\item For that particular site, compute the bosonic and fermionic Weiss fields $\mathcal{U}$ and $\mathcal{G}$ of the EDMFT impurity problem:
\begin{itemize}
\item $\mathcal{U}^{i}=W_{\mathrm{loc}}^{i}\left[1+\Pi^{\mathrm{EDMFT},i}W_{\mathrm{loc}}^{i}\right]^{-1}$,
\item $\mathcal{G}^{i}=\left[\left(G_{\mathrm{loc}}^{i}\right)^{-1}-\Sigma^{\mathrm{EDMFT},i}\right]^{-1}$.
\end{itemize}
These two fields are then rotated to the crystal-field basis:
\begin{itemize}
\item $\mathcal{U}^{i}\overset{\Theta^{i}}{\longrightarrow}\tilde{\mathcal{U}}^{i}$,
\item $\tilde{\mathcal{G}}^{i}=\left(\Theta^{i}\right)^{T}\mathcal{G}^{i}\Theta^{i}$.
\end{itemize}
\item A continuous-time Monte Carlo impurity solver\cite{Werner2006,Hafermann2013} for models with dynamically screened interactions\cite{Werner2010} provides  the density-density correlator $\tilde \chi_\text{imp}$ and the impurity Green's function $\tilde{G}_{\mathrm{imp}}$ in the crystal field basis ($0\le \tau\le \beta$)
\begin{itemize}
\item $\tilde{\chi}_{\mathrm{imp}}^{i}=\left\langle \tilde{\hat{n}}_{\alpha}\left(\tau\right)\tilde{\hat{n}}_{\beta}\left(0\right)\right\rangle $,
\item $\tilde{G}_{\mathrm{imp}}^{i}$,
\end{itemize}
which are used to solve two Dyson equations, and to extract the local EDMFT polarization and self-energy:
\begin{itemize}
\item $\tilde{\Pi}^{\mathrm{EDMFT,}i}=\tilde{\chi}_{\mathrm{imp}}^{i}\left[\tilde{\mathcal{U}}^{i}\tilde{\chi}_{\mathrm{imp}}^{i}-1\right]^{-1}$,
\item $\tilde{\Sigma}^{\mathrm{EDMFT,}i}=\left(\tilde{\mathcal{G}}^{i}\right)^{-1}-\left(\tilde{G}_{\mathrm{imp}}^{i}\right)^{-1}$.
\end{itemize}
\item The updated polarization and self-energy in the crystal-field basis is then transformed back to the orbital basis for each site using the corresponding site-dependent rotation. This generates a set of $N_{s}$ different pairs:
\begin{itemize}
\item $\tilde{\Pi}^{\mathrm{EDMFT,}i}\overset{\left(\Theta^{i}\right)^{T}}{\longrightarrow}\Pi^{\mathrm{EDMFT,}i}$,
\item $\Sigma^{\mathrm{EDMFT,}i}=\Theta^{i}\tilde{\Sigma}^{\mathrm{EDMFT,}i}\left(\Theta^{i}\right)^{T}$.
\end{itemize}
The site-diagonal $\Pi^{\mathrm{EDMFT}}$ and $\Sigma^{\mathrm{EDMFT}}$ are then substituted back into step 1, and the loop is iterated until a converged solution if obtained. 
\end{enumerate}
It is worth mentioning that, for LDA inputs with degenerate orbitals, if the invariance conditions for $U$ and $J$ are fulfilled, the rotation of $\mathcal{U}$ can be avoided. This is however not the case for $\Pi$, since the charge susceptibility will reflect the occupations of the realigned fermionic levels, thus preventing any assumption on SO$\left(3\right)$ invariance. 
%
%
%
%
\section{Results}\label{section:Results}
\subsection{Interaction strengths and electronic properties of the two structures}\label{subsection:properties}

We now apply the above {\it ab-initio} $GW$+EDMFT scheme to investigate the electronic structure of Ca$_2$RuO$_4$. This is an interesting test case for our scheme, since Ca$_2$RuO$_4$ exhibits a metal-insulator transition that is not linked to magnetic ordering. Here, we should note that spin correlations are not expected to be accurately described by any mean-field-based theory, and in the case of Ca$_2$RuO$_4$ they will be influenced by SOC. We therefore do not consider magnetic ordering in the present study. In fact, the structural deformation from the $L$-Pbca to the $S$-Pbca structure at 360~K, which shortens the $\mathbf{c}$ axis, appears to be the main driving force of the transition from a metallic high-temperature to a Mott insulating low-temperature state.\cite{Gorelov2010} The experimental gap in the $S$-Pbca structure is $\sim0.2$~eV. 

To investigate the transition, we performed two separate calculations, one for each structure, at temperatures where the corresponding electronic phases are experimentally found. Specifically, we considered  $T^{S\text{-Pbca}}=290$~K and $T^{L\text{-Pbca}}=580$~K, using a Matsubara frequency cutoff at $150$~eV for both the fermionic and bosonic fields. We considered all the four sites in the lattice problem and the corresponding $GW$ contributions, but only a single impurity problem for the calculation of the local self-energy and polarization, using the method described in Sec.~\ref{subsection:Real_space_GW+EDMFT}. The site-dependent rotations to the crystal-field basis yield a level alignment in agreement with Ref.~\onlinecite{Gorelov2010}, with a splitting between the lowest and the middle levels decreasing from $0.3$~eV to $0.1$~eV as one switches from the $S$ to the $L$ structure:

\begin{equation}
\varepsilon_{\mathrm{loc}}^{S\text{-Pbca}}{=}\left(\begin{array}{c|c}
\left|1\right\rangle  & -0.386\\
\left|2\right\rangle  & -0.084\\
\left|3\right\rangle  & -0.072
\end{array}\right),\;
\varepsilon_{\mathrm{loc}}^{L\text{-Pbca}}{=}\left(\begin{array}{c|c}
\left|1\right\rangle  & -0.321\\
\left|2\right\rangle  & -0.216\\
\left|3\right\rangle  & -0.206
\end{array}\right).
\label{eqn:splitting} 
\end{equation}
Even though we perform the tensor analogue of this operation on $\mathcal{U}$ and $\Pi$ at all the Matsubara points, we report here only the effect of the basis change on the local $U_{\mathrm{cRPA}}$ interaction at $\omega=0$:
\begingroup 
\setlength\arraycolsep{1.5pt} 
\begin{equation}
U_{\mathrm{cRPA}}^{S\text{-Pbca}}\left(0\right){=}\begin{pmatrix}
 d_{xz} & d_{yz} & d_{xy}\\
\hline 2.55 & 1.94 & 1.76 \\
\cline{1-1} \multicolumn{1}{c|}{0.30} & 2.49 & 1.74 \\
\cline{2-2} 0.27 & \multicolumn{1}{c|}{0.27} & 2.34
\end{pmatrix}
\xrightarrow{\makebox[8pt]{\small$\Theta^{i}$}}
\begin{pmatrix}
\left|1\right\rangle  & \left|2\right\rangle  & \left|3\right\rangle \\
\hline 2.34 & 1.74 & 1.75 \\
\cline{1-1} \multicolumn{1}{c|}{0.27} & 2.51 & 1.93 \\
\cline{2-2} 0.27 & \multicolumn{1}{c|}{0.29} & 2.54 
\end{pmatrix},
\label{eqn:Urpa_S}
\end{equation}
\begin{equation}
U_{\mathrm{cRPA}}^{L\text{-Pbca}}\left(0\right){=}\begin{pmatrix}
 d_{xz} & d_{yz} & d_{xy}\\
\hline 2.46 & 1.87 & 1.72 \\
\cline{1-1} \multicolumn{1}{c|}{0.31} & 2.43 & 1.72 \\
\cline{2-2} 0.27 & \multicolumn{1}{c|}{0.27} & 2.30
\end{pmatrix}
\xrightarrow{\makebox[8pt]{\small$\Theta^{i}$}}
\begin{pmatrix}
\left|1\right\rangle  & \left|2\right\rangle  & \left|3\right\rangle \\
\hline 2.30 & 1.72 & 1.73 \\
\cline{1-1} \multicolumn{1}{c|}{0.28} & 2.45 & 1.85 \\
\cline{2-2} 0.28 & \multicolumn{1}{c|}{0.30} & 2.46
\end{pmatrix},
\label{eqn:Urpa_L} 
\end{equation}
\endgroup
where the upper triangular section corresponds to the density-density interactions, while the lower one gives the Hund couplings between the different orbitals. In the crystal-field basis the orbital with the lowest local energy ($\left|1\right\rangle $) has the most screened value of the interaction, while the other two have more similar values. In the $L$-Pbca case the difference is less pronounced, because of less charge reshuffling introduced by the rotation. These static values are in agreement with what is usually employed in static DFT+DMFT simulations of Ca$_2$RuO$_4$.\cite{Gorelov2010,Sutter2017,Ricco2018,Hao2020} In the present case, however, these interactions represent an upper bound for the static interaction strength, since $GW$+EDMFT takes into account additional non-local screening processes which will lead to a reduced effective impurity interaction. On the other hand, $GW$+EDMFT treats the full frequency-dependence of the interactions up to the considerably larger bare values (see below). 
\begin{figure}
\begin{center}
\includegraphics[width=0.42\textwidth]{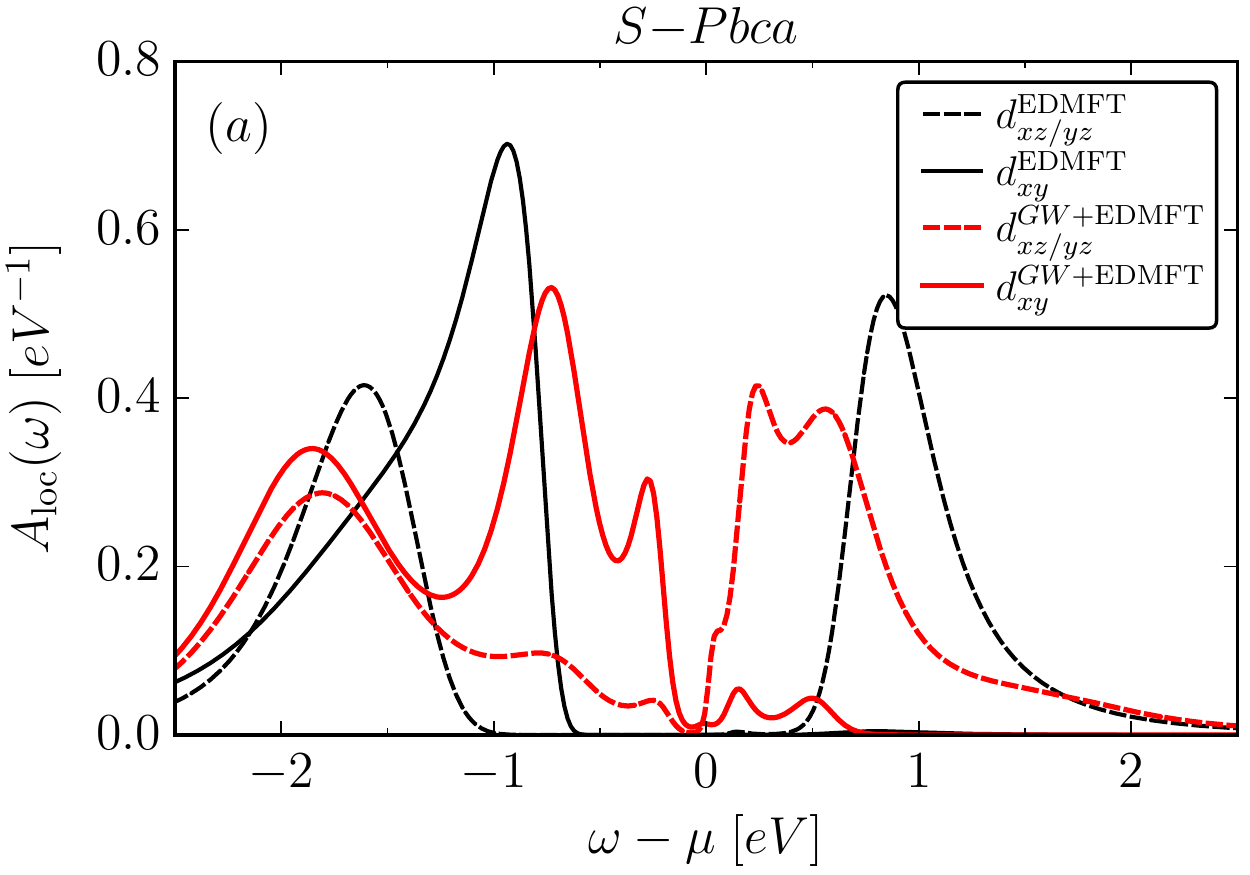}
\vspace{0.4cm} \\
\includegraphics[width=0.42\textwidth]{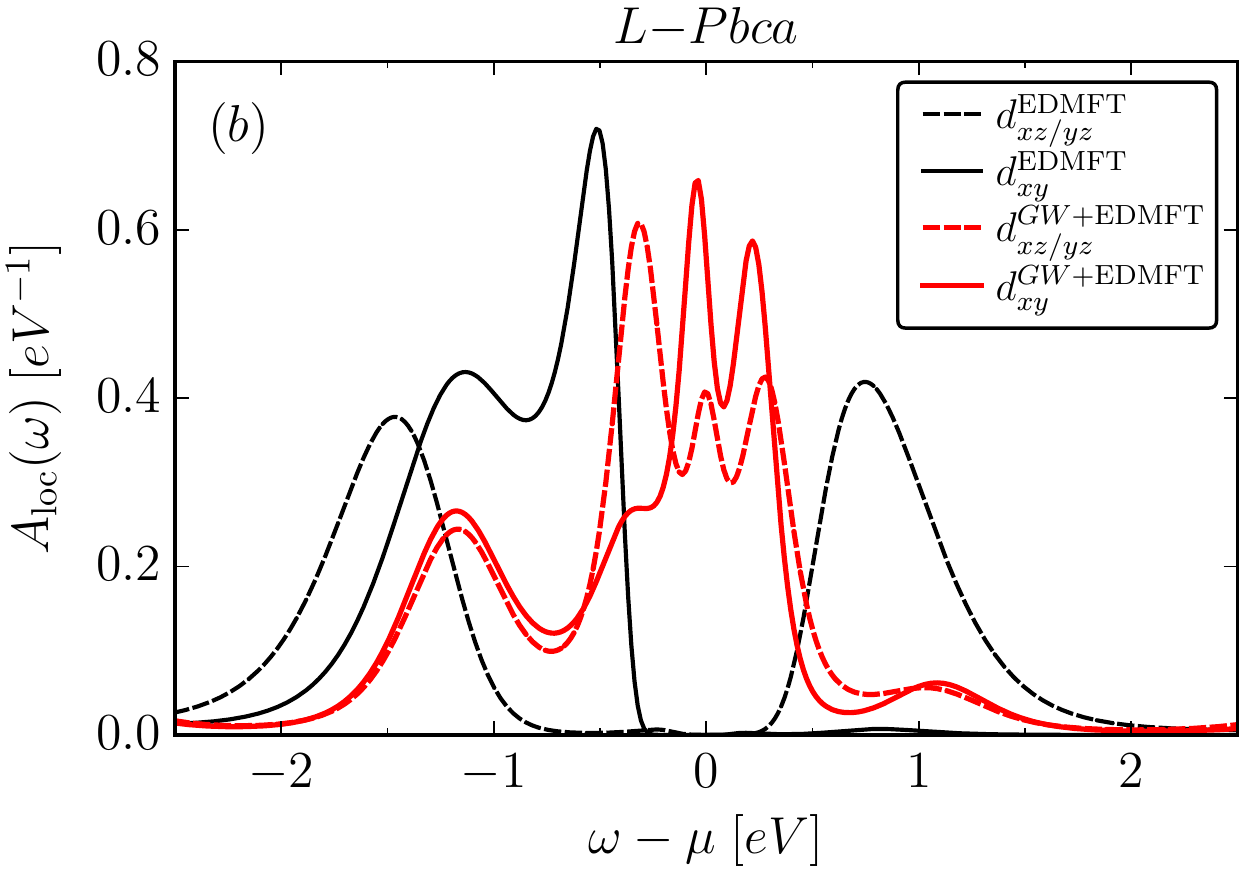}
\caption{Local spectral functions in the orbital basis. Panel (a) shows the comparison between $GW$+EDMFT (red lines) and EDMFT (black lines) in the $S$-Pbca crystal structure, while panel (b) shows the corresponding results for the $L$-Pbca structure. Solid lines refer to the $d_{xy}$ orbital, while dashed lines show the average of the $d_{xz}$ and  $d_{yz}$ spectra.\label{Localspectra}}
\end{center}
\end{figure}

A few tenths of iterations lead to a converged insulating solution for the $S$-Pbca structure and a converged metallic one for the $L$-Pbca structure, in agreement with experiment. Both are found to be stable against statistical noise and both persist in a very large temperature window: calculations between $T=250K$ and $T=1160K$ for both systems did not yield any metal-insulator transition.  This might seem in contrast with DFT+DMFT results reporting a (bad) metal state for the $S$-Pbca structure at $T=580K$.\cite{Gorelov2010} 
However, it should be noted that even if the static interactions in $GW$+EDMFT are screened down to few eV, the correlation strength is intrinsically higher than in a system with a corresponding frequency-independent interaction.\cite{Werner2016} In either case, the absence of a temperature-dependent transition in a large temperature interval around 360~K (for a given crystal structure) supports the scenario where the structural transition triggers the electronic one. Similar conclusions were drawn in a very recent study of the interplay between lattice and electronic degrees of freedom in Ca$_2$RuO$_4$.\cite{georgescu2021} 

In Fig.~\ref{Localspectra} we report the local $t_{2g}$ spectral functions in the two setups. The  charge configuration of the insulator found in the $S$-Pbca crystal structure corresponds to a nearly fully occupied $d_{xy}$ orbital, with $n_{xy}=0.97$, and two almost half-filled $d_{xz/yz}$ orbitals with $n_{xz}=0.51$ and $n_{yz}=0.52$, respectively. All the bands are insulating and the gap, measured as the distance between the two frequency points where the spectral weight vanishes, is $\sim0.2$~eV wide, in agreement with experiment. The spectral weight associated with the lower Hubbard band (LHB), of mainly $d_{xz/yz}$ character, is centered at $-1.8$~eV, while the upper Hubbard band (UHB) is centered at $0.5$~eV. The almost filled $d_{xy}$ band displays two prominent features separated by roughly $1.2$~eV. 

The metallic state obtained for the $L$-Pbca structure has negligible orbital polarization given by $n_{xz}=n_{yz}=0.67$ and $n_{xy}=0.66$, and all the bands contribute to the spectral weight at the Fermi level. Comparing the $GW$+EDMFT spectral functions with the LDA ones (see rightmost panel of Fig.~\ref{fig:Akw_G_L} below) and considering the value of the effective local interaction (Eq.~\eqref{eqn:LInt}) of $\sim1.4$-$1.6$~eV, we deduce that the three peak structure appearing in the $-0.5$~eV to $0.5$~eV energy window does not correspond to preformed Hubbard bands. This is rather a consequence of a renormalization of the LDA bands, which exhibit similar structures in the local spectral function. Also, by comparison with the LDA spectra one can see that the weight at $\sim-1.2$~eV for the $d_{xz/yz}$ orbitals is due to interactions and, as we will discuss later, may be interpreted as a LHB. 

To assess the role of the non-local screening from the $GW$ contribution, we also performed EDMFT calculations for both setups. This corresponds to setting $\Pi_{\mathbf{q}}^{GG}=0$ and $\Sigma_{\mathbf{k}}^{GW}=0$ at each iteration in the selfconsistency loop described in Sec.~\ref{subsection:Real_space_GW+EDMFT}. As can be seen from the black lines in Fig.~\ref{Localspectra}, these EDMFT results are inconsistent with experiment, since they yield a wide-gap Mott insulating state with full orbital polarization ($n_{xz}=n_{yz}=0.5$ and $n_{xy}=1$ configuration) for both structures. This further illustrates the fact that in a calculation with frequency-dependent interactions, the correlation strength is significantly stronger than what one might expect by looking at the $\omega=0$ value. The result also shows that for a meaningful self-consistent {\it ab-initio} description of the compound, one needs to take into account the non-local screening originating from the non-local interactions. In addition, the band widening effect of the $GW$ self-energy, coming mainly from the Fock term,\cite{Ayral2017} plays an important role in setting the correct balance between the bandwidth and correlation strength. 
 
\begin{figure}
\begin{center}
\includegraphics[width=0.48\textwidth]{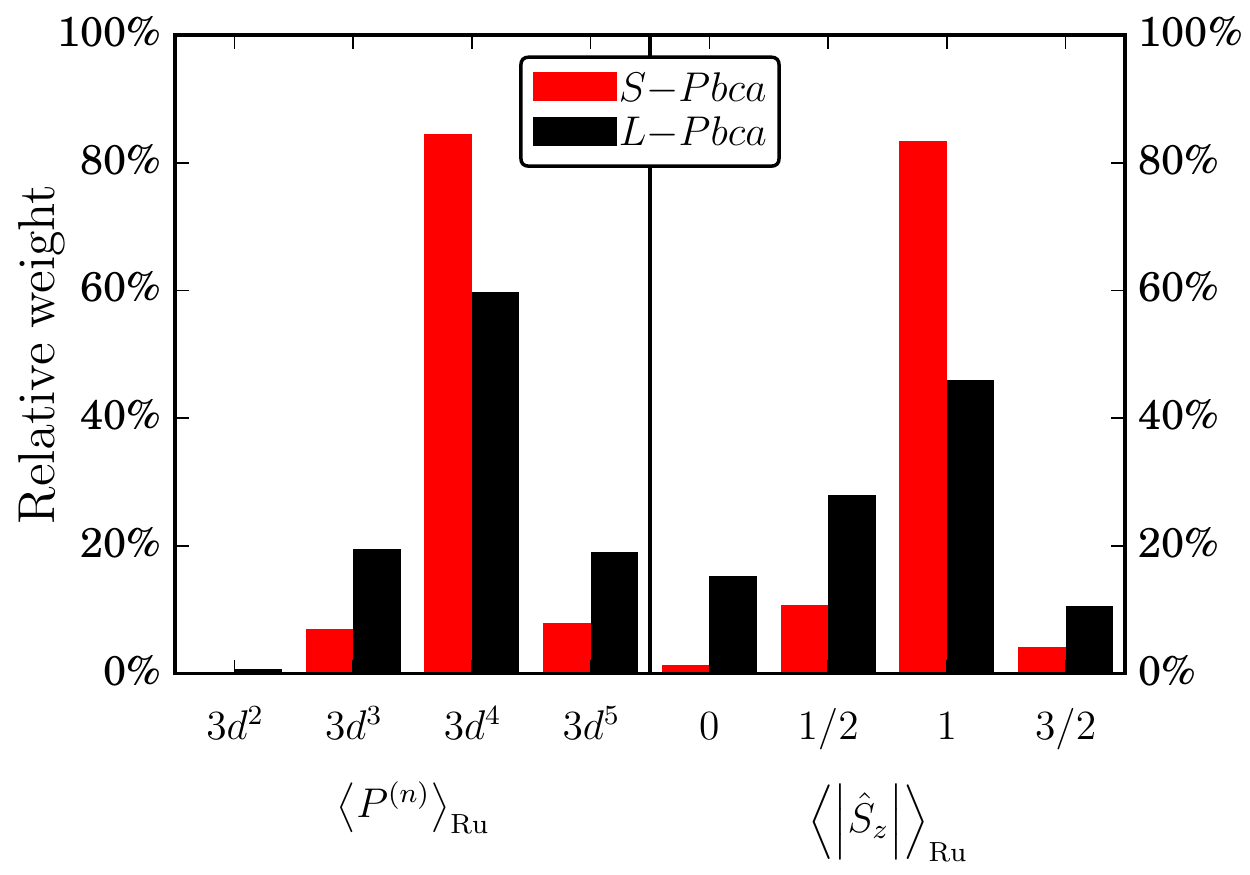}
\caption{Occupation and spin statistics for the Ru atoms. Left panel: probability distributions for the atom to be in the indicated charge states. Right panel: probability distributions for the absolute value of the $z$-component of the spin.\label{histogram}}
\end{center}
\end{figure}

Information on the local states of the correlated sites can be obtained by measuring the probability with which the Monte Carlo solver generates a given level/spin configuration. The histogram of the occupation and spin statistics is reported in Fig.~\ref{histogram} and clearly shows that the Ru atoms in the insulating phase ($S$-Pbca structure) are in a high spin $3d^4$ configuration with suppressed charge fluctuations to the $3d^3$ and $3d^5$ configurations. This is the result of a strong Hund's coupling in Ca$_2$RuO$_4$ which favors spin-1 configurations with four electrons in the $t_{2g}$ orbitals, as directly confirmed in the right panel of the figure. Both the charge and the spin are fluctuating more strongly in the metallic phase ($L$-Pbca structure), but there is still a dominant spin-1 state and a dominant $3d^4$ configuration. The high-temperature state of Ca$_2$RuO$_4$ may thus be characterized as a Hund metal.\cite{Georges2013} 
\subsection{Results for the $S$-Pbca structure}
In this section, we will analyze in more detail the properties of the low-temperature $S$-Pbca structure, and compare the $GW$+EDMFT results to experimental data.  

\subsubsection{Self-energies and effective interactions}
\begin{figure}
\begin{center}
\includegraphics[width=0.49\textwidth]{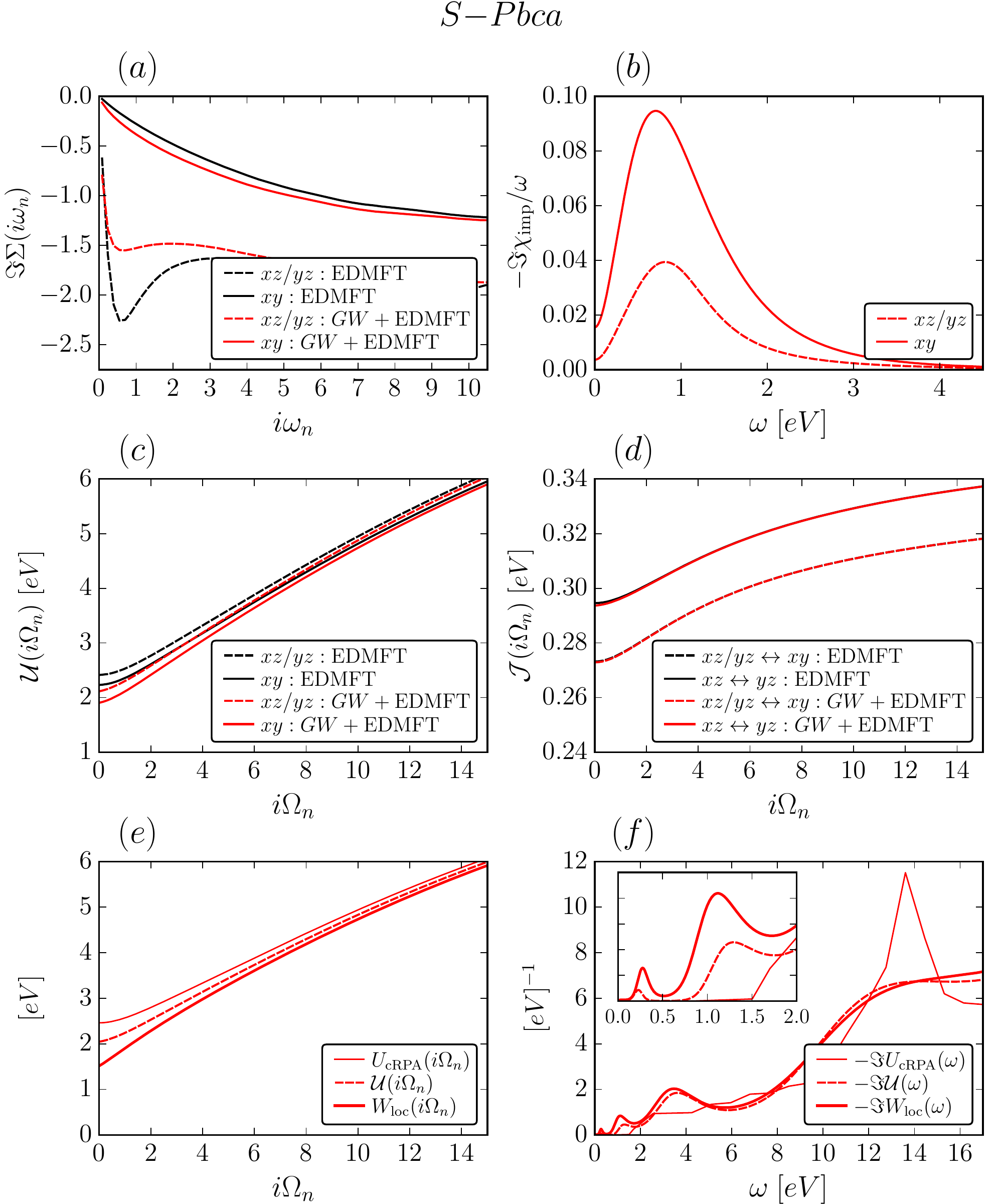}
\caption{Frequency-dependent local interactions and related quantities for the $S$-Pbca crystal structure. The red (black) lines show the $GW$+EDMFT (EDMFT) results. (a) Imaginary part of the local self-energy on the Matsubara axis, with thick (dashed) lines showing the results for the $d_{xy}$ ($d_{xz/yz}$) orbitals. (b) Strength of the local charge fluctuations in the two channels on the real frequency axis. (c,d) Effective density-density interaction $\mathcal{U}$ and Hund coupling $\mathcal{J}$ on the Matsubara axis. (e) Frequency dependence of the orbitally-averaged density-density elements of the interaction. The thin line shows the $U_{\mathrm{cRPA}}$ input (``bare" interaction), the dashed line is the local effective interaction $\mathcal{U}$ and the thick line is the fully screened interaction $W$. (f) Local bosonic spectral functions showing the energies of single-particle and collective charge excitations. In the inset the low-energy screening modes are magnified. \label{Interaction_Scell}}
\end{center}
\end{figure}

The self-consistently determined bosonic fields in $GW$+EDMFT allow us to address the screening induced by long-range interactions and to explore how dynamical screening affects the electronic structure. In Fig.~\ref{Interaction_Scell}(a) we illustrate the effect of the non-local screening by comparing the imaginary parts of the $GW$+EDMFT and EDMFT local self-energies $\Im\Sigma$ for the insulating structure. The small imaginary self-energy associated with $d_{xy}$ reflects the fact that this orbital is in an almost band insulating state, while the larger $\Im\Sigma$ for the $d_{xz/yz}$ orbitals is responsible for the gapped spectra. The absence of non-local screening in EDMFT increases the orbital polarization by almost completely filling the $d_{xy}$ orbital, so that the corresponding self-energy is further reduced. The opposite effect occurs in the $d_{xz/yz}$ orbitals which, being exactly half-filled, have a larger  $\Im\Sigma$ compared to the $GW$+EDMFT counterpart. A consistent picture is obtained from the intensity of the local charge susceptibility $-\Im\chi_{\mathrm{imp}} / \omega$ which is shown in Fig.~\ref{Interaction_Scell}(b) on the real axis. $-\Im\chi_{\mathrm{imp}} / \omega$ has a peak at an energy comparable with the splitting between the upper and lower bands. The overall magnitude is already relatively small in $GW$+EDMFT and several orders of magnitude smaller in EDMFT (not shown in the plot), where charge fluctuations are essentially frozen. 

One consequence of ignoring non-local charge fluctuations, reported in panel (c), is that the EDMFT local effective interaction tensor $\mathcal{U}$ is essentially identical to the bare $U_{\mathrm{cRPA}}$ one (compare to panel (e)), while in $GW$+EDMFT, even in the insulating state, it is significantly reduced for both the $d_{xz/yz}$ and $d_{xy}$ orbitals. The effective Hund coupling $\mathcal{J}$, reported in Fig.~\ref{Interaction_Scell}(d), is almost unchanged from the cRPA result in both methods, and it has little frequency dependence ($\sim 15\%$ difference between the bare and screened value), which shows that screening has little effects on this quantity. The situation is different for the effective impurity model interaction tensor, whose static value in the orbital basis reads
\begingroup
\setlength\arraycolsep{1.5pt}
\begin{equation}
\mathcal{U}^{S\text{-Pbca}}\left(0\right){=}\begin{pmatrix}
 d_{xz} & d_{yz} & d_{xy}\\
\hline 2.08 & 1.49 & 1.29 \\
\cline{1-1} \multicolumn{1}{c|}{0.29} & 2.07 & 1.29 \\
\cline{2-2} 0.27 & \multicolumn{1}{c|}{0.27} & 1.85
\end{pmatrix}.
\label{eqn:SInt} 
\end{equation}
\endgroup
By comparing Eq.~\eqref{eqn:SInt} to Eq.~\eqref{eqn:Urpa_S} one can see that the screening affects the three orbitals significantly.

In the last two panels of Fig.~\ref{Interaction_Scell} we illustrate how the average of the diagonal components of the $U_{\mathrm{cRPA}}$ tensor are modified by the screening on the Matsubara and real frequency axis. Given the insulating state in the $S$-Pbca structure one may not expect large effects on the effective interaction strength, but $\mathcal{U}(\omega=0)$ is nevertheless reduced relative to $U_\text{cRPA}(\omega=0)$ by nearly half an eV. The imaginary part of the same quantities, plotted on the real-frequency axis, reveals the characteristic energies of screening modes associated with (local and nonlocal) charge excitations in the low-energy space, which are excluded from the cRPA input. From Fig.~\ref{Interaction_Scell}(f) one sees that the bosonic spectral functions, while following the cRPA behavior at high energies, exhibit two additional modes at $0.2$~eV and $1.1$~eV in $-\Im W_{\mathrm{loc}}(\omega)$, and similarly also for $\mathcal{U}$. We associate the peak at $0.2$~eV with charge excitations across the gap, which has a similar magnitude (Fig.~\ref{Localspectra}(a)). The second peak is located at an energy comparable to the separation between the main features with $d_{xy}$ and $d_{xz/yz}$ character in the occupied and unoccupied parts of the spectra. A possible explanation for these features involves local multiplet excitations with magnitude $\sim3\mathcal{J}$, 
as proposed in Ref.~\onlinecite{Sutter2017}.  
%
\begin{figure*}[ht]
\includegraphics[width=0.9\textwidth]{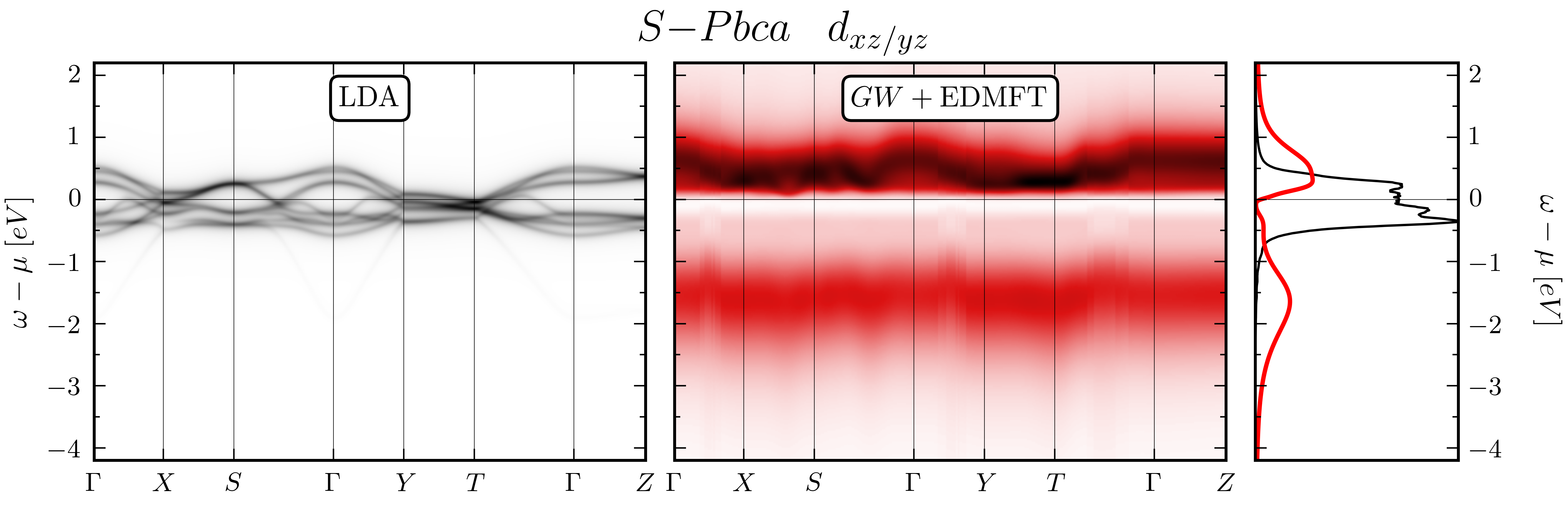}
\vspace{0.2cm} \\
\includegraphics[width=0.9\textwidth]{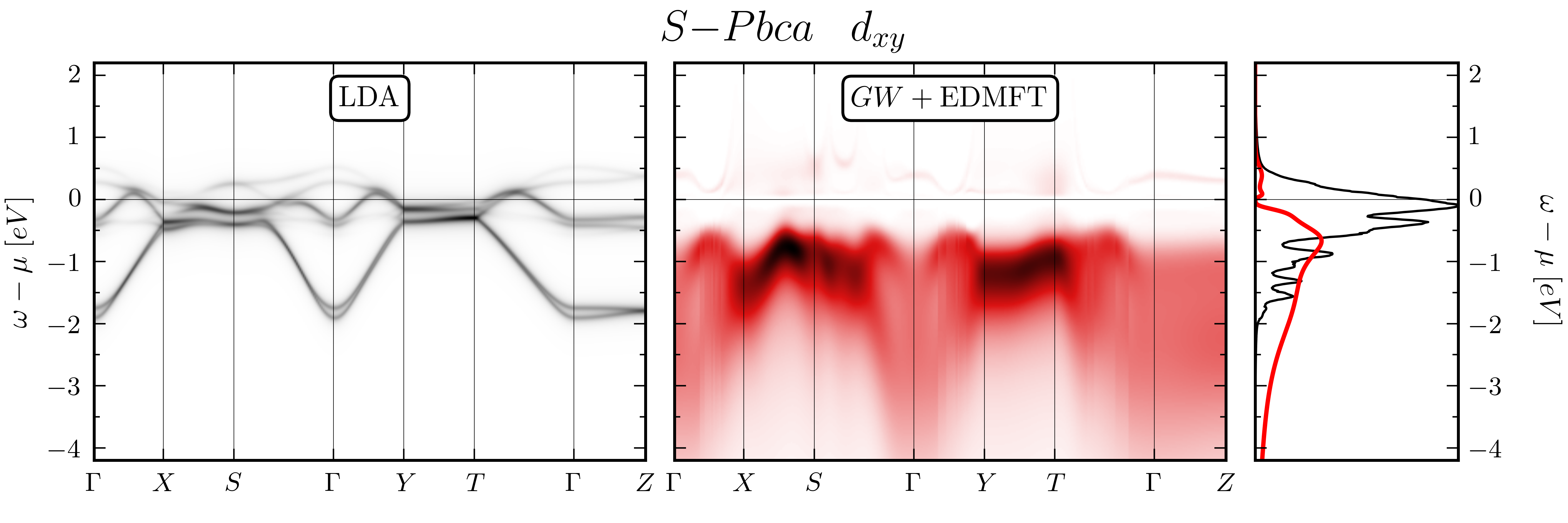}
\caption{Momentum-resolved spectral function for Ca$_{2}$RuO$_{4}$ in the $S$-Pbca crystal structure at inverse temperature $\beta=40$ eV$^{-1}$ obtained by applying the maximum entropy method to the lattice Green's function computed along the indicated path. The results are provided in the basis with well-defined orbital character. Dark colors correspond to high intensities. The top panels show the (average) result for  the $d_{xz/yz}$ orbitals, while the bottom panels show the results for the $d_{xy}$ orbital. The left panels plot the LDA bandstructure, the middle panels the correlated electronic structure from $GW$+EDMFT, and the right panels the local spectra (black for LDA and red for $GW$+EDMFT). 
\label{fig:Akw_G_S}}
\end{figure*}
\subsubsection{Momentum-resolved spectra and comparison with photoemission experiments}
The momentum-resolved spectral function for the $S$-Pbca structure is shown in the middle panel of Fig.~\ref{fig:Akw_G_S} for a path along high symmetry lines in the first Brillouin zone, as illustrated in Fig.~\ref{Figure: Structure}(b). These results confirm the insulating state of the system and the positions of the (weakly dispersive) Hubbard bands deduced from the local spectral functions. They are also in good agreement with the previous results obtained with the DFT+DMFT method,\cite{Gorelov2010,Sutter2017,Ricco2018} although one needs to keep in mind that the latter approach involves several adjustable parameters. 
The comparison with the momentum-resolved LDA spectral functions in the orbital basis (left panels) reveals how the  dispersive $d_{xz/yz}$ bands are split into strongly broadened Hubbard bands, while the $d_{xy}$ band is shifted to lower energies and also substantially broadened. In the latter case, however, the broadening may be to a large extent due to the limitations of maximum entropy analytical continuation,\cite{Jarrell1996} since (as shown in Fig.~\ref{Interaction_Scell}(a)) the self-energy effects in the almost completely filled $d_{xy}$ orbital are weak. 

\begin{figure}
\begin{centering}
\includegraphics[width=0.42\textwidth]{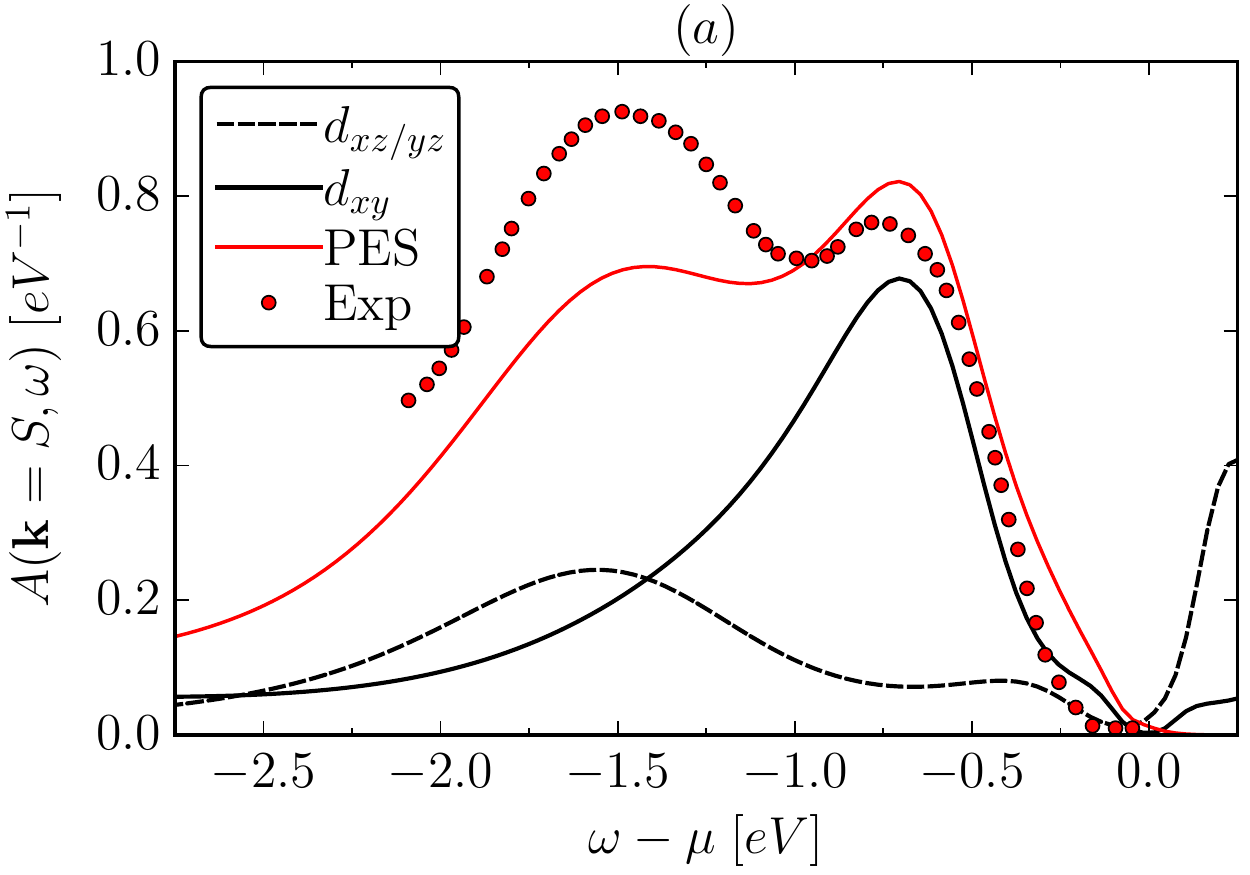}
\vspace{0.4cm} \\
\includegraphics[width=0.42\textwidth]{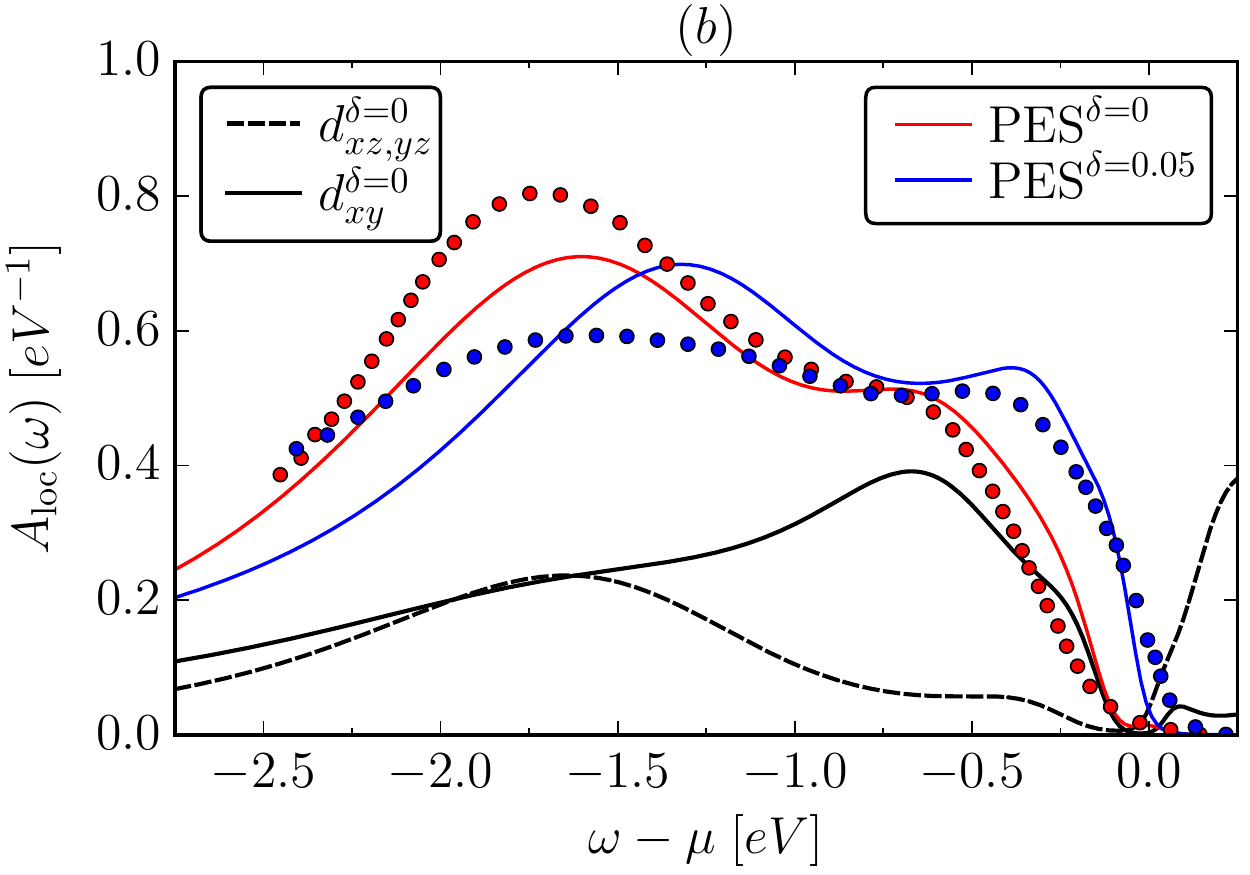}
\par\end{centering}
\caption{Comparison between theoretical bulk calculations and experimental spectra in the $S$-Pbca crystal structure. (a) Results at the $S$ point. Black thick (dashed) lines show the $d_{xy}$ ($d_{xz/yz}$) spectral functions, the red line is the theoretical PES profile, while the red dots are experimental data extracted from Fig.~2(c) in Ref.~\onlinecite{Sutter2017}. (b) Local spectral functions. Black thick (dashed) lines indicate the $d_{xy}$ ($d_{xz/yz}$) local spectral function, the red and blue lines are the theoretical PES profile for the undoped and doped system respectively, while the red and blue dots are extracted from the PES reported in Fig.~2(c) of Ref.~\onlinecite{Miyashita2021} (same color code). The theoretical PES profile is obtained by multiplying the spectral function by the Fermi distribution for $\beta=40$ eV$^{-1}$.\label{Spoint}}
\end{figure}

ARPES experiments on the $S$-Pbca structure\cite{Sutter2017} report the presence of two main peaks in the occupied part of the spectrum. These can be identified with the LHB and associated with the $d_{xz/yz}$ orbitals, but may also contain  intensity coming from the $d_{xy}$ band. Their position at the $S$ point is measured at $-1.7$~eV and $-0.8$~eV respectively. In Fig.~\ref{Spoint}(a) we show the comparison between the photoemission spectrum (PES) computed from the spectral function at the $S$ point by multiplying with the Fermi function for temperature 290~K, and the experimental data extracted from Fig.~2(c) of Ref.~\onlinecite{Sutter2017}. There is a good agreement between the theoretical and experimental results concerning both the position of the peaks and their orbital character, while there is some mismatch in the relative weight of the peaks. This may have several origins: i) matrix element effects, ii) the difficulties of maximum entropy analytical continuation in resolving high-energy spectral features, and ii) the omission of oxygen $2p$ bands, located below $-3$~eV, whose tails could have an effect on the signal up to $-1.7$~eV. 

As a further test of our $GW$+EDMFT scheme, we check the {\it ab-initio} results against recent angle-integrated photoemission experiments,\cite{Miyashita2021} which analyzed how the position of the peaks in the spectra evolve with hole doping. Experimentally, it has been found that, by increasing doping, a spectral weight redistribution occurs at low energy and a metallic state emerges on the sample surface, whereas the bulk remains insulating.\cite{Miyashita2021} In the simulations, we considered a homogeneous system with a doping of $\delta=0.05$, corresponding to a total density of 15.8 electrons within the four-site unit cell. In principle this would require us to perform a separate $G_0W_0$ and cRPA calculation for the low-energy model of the non-stoichiometric system, but, because of the small hole concentration, we used the previous undoped input for simplicity.

In Fig.~\ref{Spoint}(b) we report the comparison between our bulk calculations and the experimental PES taken from Fig.~2(c) of Ref.~\onlinecite{Miyashita2021}. The results for the stoichiometric compound are in remarkably good agreement with the experimentally measured local PES concerning both the positions and relative weights of the peaks. On the other hand, the doped setup only reproduces the measured spectral weight transfer at low energy, while the shift of the higher energy peak (thick blue line), which is due to the insulating $d_{xz/yz}$ orbitals, is less prominent in the experimental data. Whether or not this is related to the insulating bulk in the experimentally investigated compound is an interesting open question. 

\subsection{Results for the $L$-Pbca structure}
In this section, we will analyze in more detail the properties of the high-temperature $L$-Pbca structure. 

\subsubsection{Self-energies and effective interaction}

\begin{figure}
\begin{center}
\includegraphics[width=0.49\textwidth]{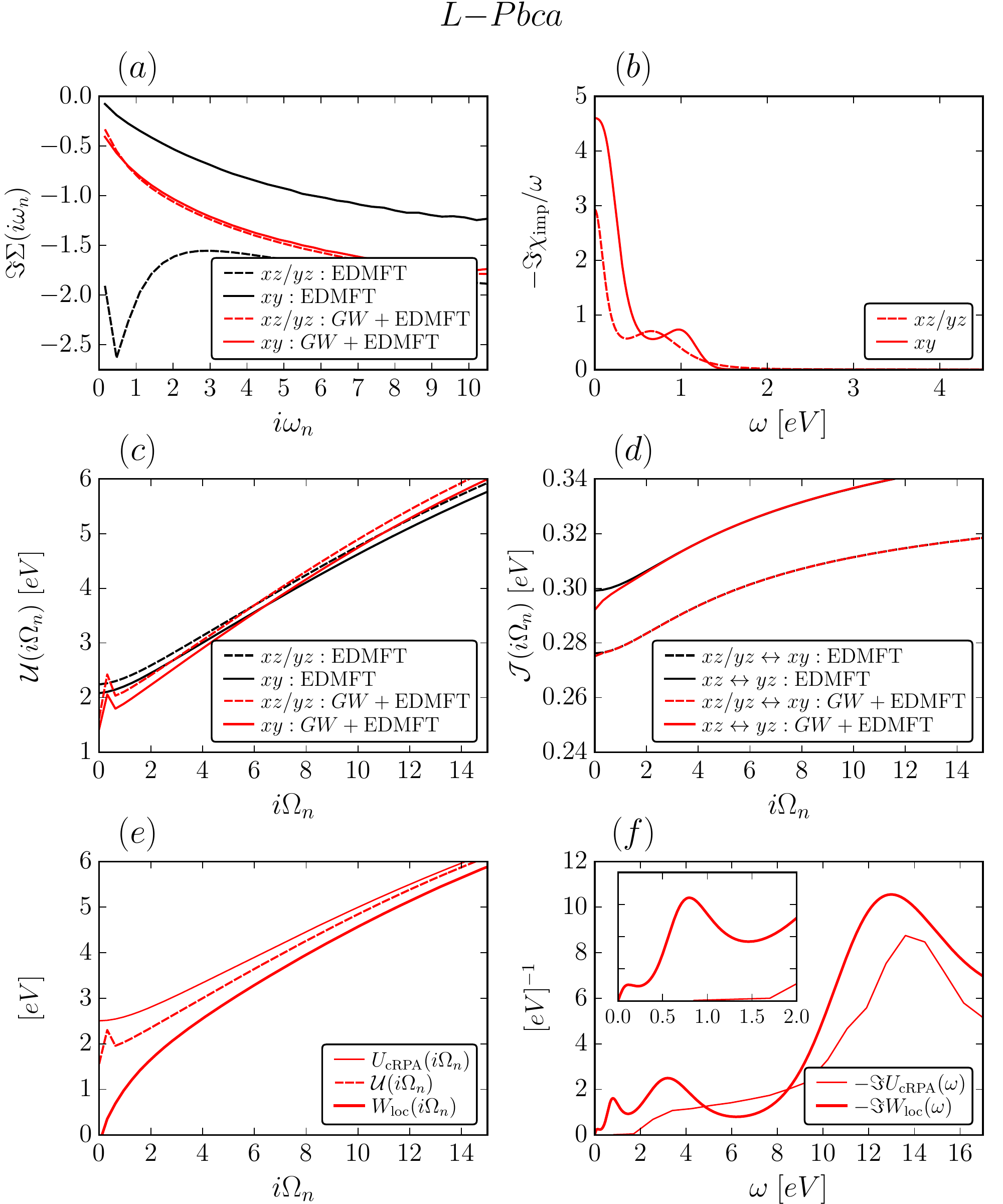}
\caption{Frequency-dependent local interactions and related quantities in the $L$-Pbca crystal structure. The red (black) lines show the results from $GW$+EDMFT (EDMFT). (a) Imaginary part of the local self-energy on the Matsubara axis. (b) Strength of the local charge fluctuations in the two channels on the real frequency axis. (c,d) Density-density interaction $\mathcal{U}$ and Hund coupling $\mathcal{J}$ on the Matsubara axis. (e) Frequency dependence of the orbitally-averaged density-density elements of the interaction. (f) Local bosonic spectral functions showing the energies of single-particle and collective charge excitations. The analytical continuation of $\mathcal{U}$ is absent due to the non-causality which prevents maximum entropy analytical continuation.\label{Interaction_Lcell}}
\end{center}
\end{figure}

\begin{figure*}[ht]
\includegraphics[width=0.9\textwidth]{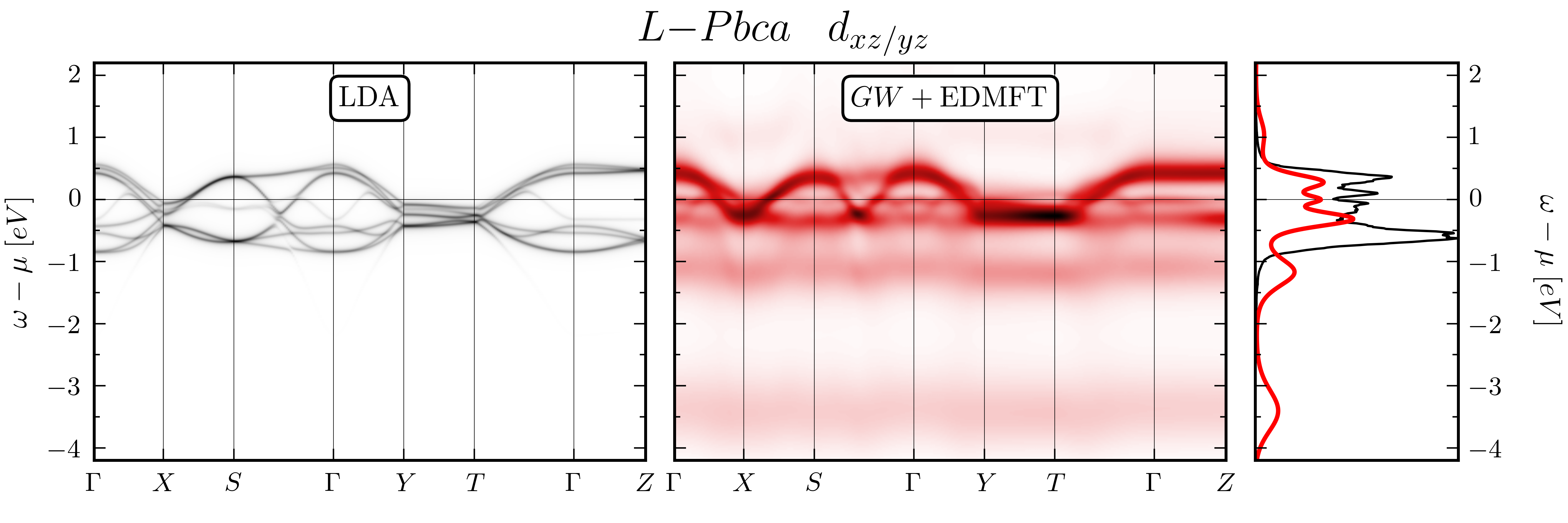}
\vspace{0.2cm} \\
\includegraphics[width=0.9\textwidth]{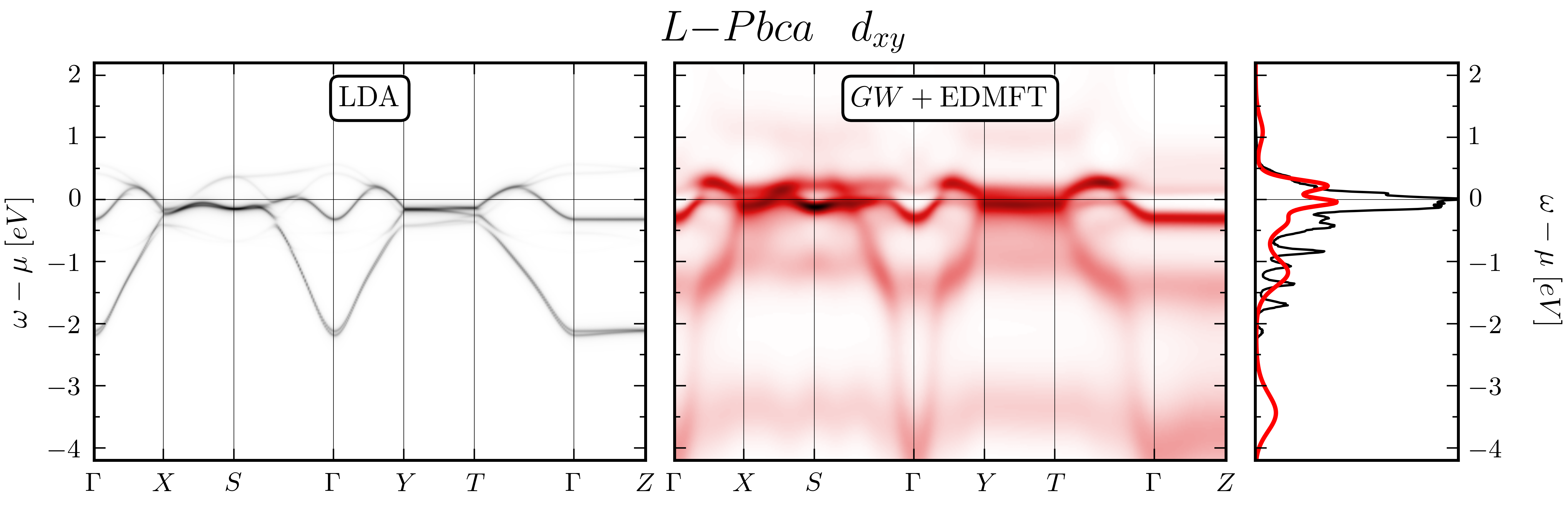}
\caption{Momentum-resolved spectral functions for Ca$_{2}$RuO$_{4}$ in the $L$-Pbca crystal structure at $\beta=20$ eV$^{-1}$ obtained by applying the maximum entropy method to the lattice Green's function computed along the indicated path in the basis where the orbital character is well defined. Dark colors correspond to high intensities. The top panels show the (average) result for  the $d_{xz/yz}$ orbitals, while the bottom panels show the results for the $d_{xy}$ orbital. The left panels show the LDA bandstructure, the middle panels the correlated electronic structure from $GW$+EDMFT, and the right panels the local spectra (black for LDA and red for $GW$+EDMFT). \label{fig:Akw_G_L}}
\end{figure*}

In the $L$-Pbca structure, $GW$+EDMFT yields the expected metallic solution, while EDMFT produces a wide-gap insulator. Therefore, the comparison between the local self-energies obtained with the two methods, reported in Fig.~\ref{Interaction_Lcell}(a), shows pronounced differences: in EDMFT the $d_{xy}$ orbital is completely filled and $\Im \Sigma$ is very small, while the self-energy is much larger for the half-filled $d_{xz/yz}$ orbitals. The opposite result is obtained in the metal, where all orbitals contribute to the spectral weight at the Fermi level. Here, all the self-energies of the $t_{2g}$ manifold are quite similar. The local charge fluctuations, illustrated in Fig.~\ref{Interaction_Lcell}(b), are several orders of magnitude larger in the metallic state than in the EDMFT case for the same structure, or for the $GW$+EDMFT solution for the $S$-Pbca setup. $-\Im\chi_{\mathrm{imp}} / \omega$ now exhibits a large peak at zero energy, with a relative strength for the different orbitals in agreement with the relative $\omega=0$ contribution of the  orbitals to the local spectral function shown in Fig.~\ref{Localspectra}(b). We also notice a second peak that roughly matches the energy separation between the highest occupied state and the lowest available one. 

In panels (c) and (d) of Fig.~\ref{Interaction_Lcell} we report the components of the local effective interaction as a function of Matsubara frequency: the EDMFT results, as in the previous structure, are very close to the corresponding cRPA results, while significant screening effects are found in both the Hund coupling and density-density interactions in the metallic case. The magnitude of the total screening in the metal solution is $0.8$-$0.9$~eV, as can be inferred by comparing the static value of the effective interaction tensor in the orbital basis,

\begingroup
\setlength\arraycolsep{1.5pt} 
\begin{equation}
\mathcal{U}^{L\text{-Pbca}}\left(0\right){=}\begin{pmatrix}
 d_{xz} & d_{yz} & d_{xy}\\
\hline 1.63 & 1.04 & 0.89 \\
\cline{1-1} \multicolumn{1}{c|}{0.29} & 1.62 & 0.88 \\
\cline{2-2} 0.26 & \multicolumn{1}{c|}{0.27} & 1.43
\end{pmatrix},
\label{eqn:LInt} 
\end{equation}
\endgroup
with the corresponding cRPA matrix of Eq.~\eqref{eqn:Urpa_L}.

The frequency dependence of $\mathcal{U}\left(i\Omega_n\right)$ reveals the presence of bosonic modes with non-causal (negative) spectral weight. This is a known feature\cite{Nilsson2017} which may occur in the auxiliary impurity problem as a result of the local approximation. Modified self-consistency equations designed to remove this kind of non-causality have recently been proposed,\cite{backes2020} but still need to be systematically investigated in model and materials contexts. Here, we use the original $GW$+EDMFT self-consistency scheme,\cite{Biermann2003,Nilsson2017} since the non-causalities in the dynamical mean fields of the auxiliary impurity problem do not produce any anomalies in physical quantities. In particular, the fully screened local interaction of the impurity and lattice system, $W_{\mathrm{loc}}(i\Omega_n)$, is causal, as illustrated in Fig.~\ref{Interaction_Lcell}(e).

The imaginary part of the analytical continuation of $W_\text{loc}$ to the real axis, $-\Im W_{\mathrm{loc}}\left(\omega\right)$, is reported in Fig.~\ref{Interaction_Lcell}(f) and reveals the presence of low-energy charge excitations at $\omega\sim0.9$~eV. In this plot we did not include $-\Im \mathcal{U}\left(\omega\right)$ because its non-causality prevents us from employing the maximum entropy method, but looking at the Matsubara axis data, we expect it to be reasonably close to the spectrum shown in Fig.~\ref{Interaction_Scell}(f).

\subsubsection{Momentum-resolved spectra and Fermi surface}
In Fig.~\ref{fig:Akw_G_L} we compare the momentum-resolved LDA spectral functions (left panels) to the interacting spectral functions for the $L$-Pbca structure from $GW$+EDMFT (middle panels). All the orbitals remain metallic, but, as mentioned in Sec.~\ref{subsection:properties}, we find a significant correlation-induced spectral weight redistribution and the appearance of satellite structures near $-1.2$~eV and $-3.5$~eV.

Near the Fermi level, the $d_{xz/yz}$ bands are significantly renormalized, especially in the occupied part of the spectrum, while the $d_{xy}$ bands are mainly broadened, but not much renormalized. A striking feature is the 
almost nondispersive band at $-3.5$~eV, which is completely absent in LDA, and beyond the energy range of the local effective interaction $\mathcal{U}$ in Eq.~\eqref{eqn:LInt}. This high energy peak most likely originates from physics involving the higher-energy degrees of freedom, encoded in the frequency dependence of the interaction parameters. The fact that screening modes between $3$~eV and $4$~eV are present in $-\Im W_{\mathrm{loc}}\left(\omega\right)$ (and likely also in $\mathcal{U}$), as shown in Fig.~\ref{Interaction_Lcell}(f), suggests that this high-energy spectral feature is a plasmon satellite. 

We next discuss the almost dispersionless band at $-1.2$~eV in the $d_{xz/yz}$ spectrum, and an analogous weaker feature in the $d_{xy}$ spectrum. While there is no spectral weight near $-1.2$~eV in the $d_{xz/yz}$ LDA spectral function, the $d_{xy}$ orbital contributes some spectral weight, which however originates from the strong dispersion in the vicinity of the $\Gamma$ point. The almost dispersionless character of this feature, and its shift relative to the unoccupied part of the spectrum, which matches the magnitude of the interactions in Eq.~\eqref{eqn:LInt}, suggests an interpretation as the LHB of the $t_{2g}$ shell. (The upper Hubbard band cannot be easily identified due to strong damping induced by the large $\Im\Sigma$ in the corresponding energy region.)

\begin{figure}
\begin{center}
\includegraphics[width=0.49\textwidth]{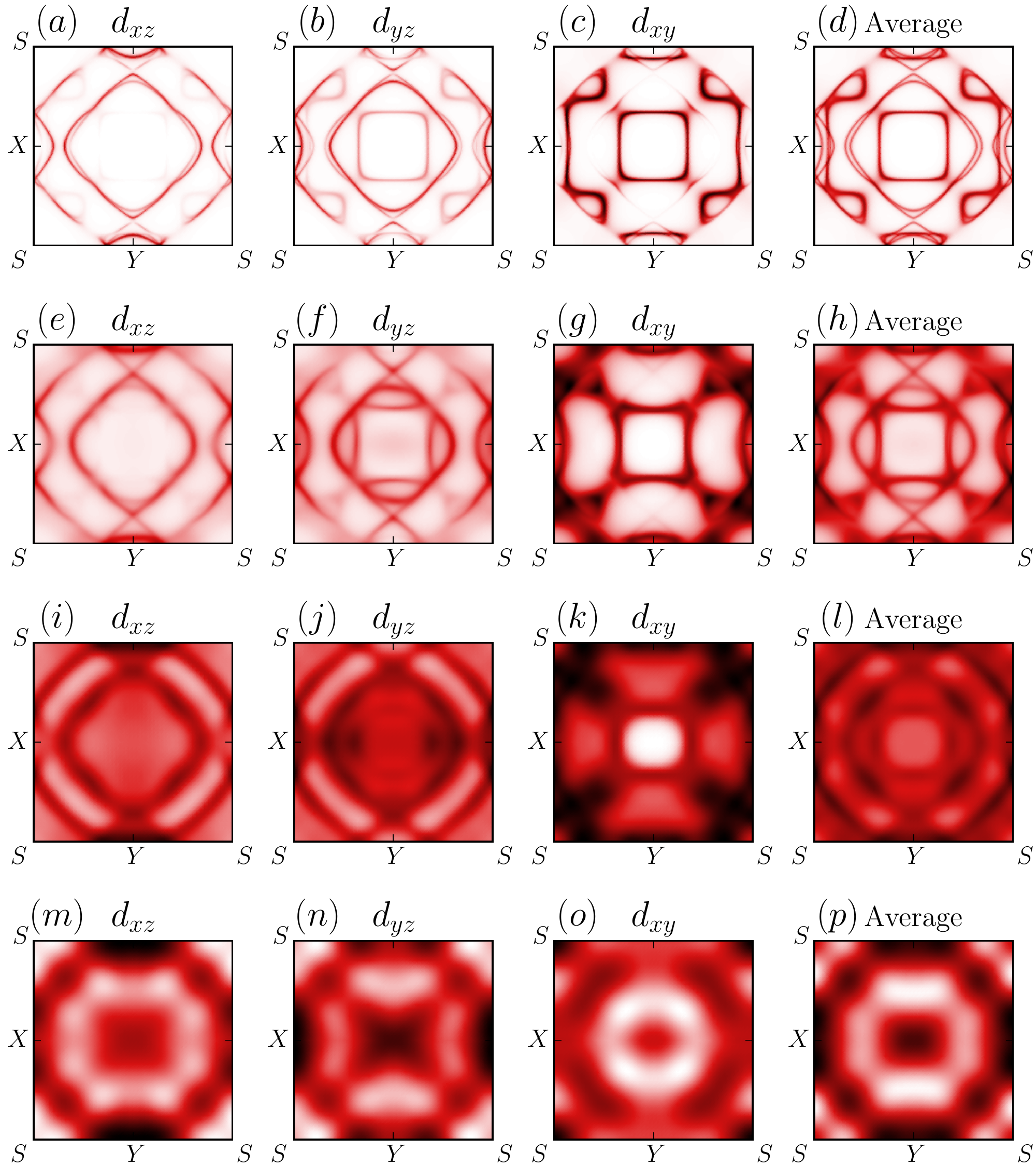}
\caption{(a-d) Orbital resolved DFT Fermi surfaces (spectral weight at $\omega=0$) of metallic Ca$_2$RuO$_4$ in the $L$-Pbca structure. (e-h) Quasiparticle Fermi surfaces obtained by correcting the single-particle Hamiltonian with the static self-energy as described in the text. (i-l) Fully interacting Fermi surfaces. (m-p) Modulation of the quasi-particle weight $Z_{\mathbf{k}}$ in the $\left\{k_x,k_y\right\}$ plane. Momentum-dependent variations of $Z_{\mathbf{k}}$ are within $6\%$ of the local value reported in the main text. In all panels dark colors correspond to high intensities.} \label{Fermi}
\end{center}
\end{figure}

The orbital resolved Fermi surfaces (FS) are plotted in Fig.~\ref{Fermi} for different methods and using different procedures: in panels (a-d) we report the FS resulting from the DFT bandstructure, in the following (e-h) panels we computed the quasiparticle FS by correcting the single-particle Hamiltonian with the 
static component of the
self-energy, with the scattering rates (diagonal imaginary parts) removed. This procedure provides an indication on the band renormalizations induced by local and non-local interactions, while suppressing the broadening due to scattering processes. In panels (i-l) the $\omega=0$ value of the momentum-resolved spectral function is reported. Our results for the FS are in agreement with the available experimental data, \cite{Ricco2018} up to small details which may be explainable by the fact that the experiments were performed on a strained and slightly doped sample. 

In panels (m-p) of the same figure we furthermore provide an estimate of the non-local correlation effects, i.e. of the $\mathbf{k}$-dependent self-energy, by plotting the quasiparticle weight 
\begin{equation}
Z_{\mathbf{k}} = \left( 1-\left.\frac{\partial\Im\Sigma\left(\mathbf{k},i\omega_{n}\right)}{\partial\omega_{n}}\right|_{0} \right)^{-1}.
\end{equation}
$Z_{\mathbf{k}}$ cannot be obtained from standard DFT+DMFT calculations, which consider only local correlation effects.  It may be used to quantify the actual correlations, which depend on the interaction parameters and on the density, since the closer the band is to an empty/filled configuration the higher $Z_{\mathbf{k}}$. From the local quasiparticle weights of the two manifolds we find that the metallic state is indeed strongly correlated, since $Z_{\mathrm{imp}}^{xz,yz}\sim0.328$ and $Z_{\mathrm{imp}}^{xy}\sim0.271$, while the momentum-dependent variations are about $6\%$ of these values. Figure~\ref{Fermi}(m-p) shows these variations as an intensity plot. One sees that the $Z_{\mathbf{k}}$ associated to different orbitals are not simply the negative of the corresponding Fermi surfaces, as one would naively expect by assuming a high $Z_{\mathbf{k}}$ wherever the spectral weight at the Fermi energy is low (e.~g. close to the $\Gamma$ point). The evolution of $Z_{\mathbf{k}}$ is indeed highly non-trivial: for instance the $d_{xy}$ behavior in the $Y$-$X$ direction is opposite to the one near the $\Gamma$ point.
%
%
%
\section{Summary and Conclusions}\label{section:Conclusions}
In this work we have tested the $GW$+EDMFT method on the experimentally well-characterized layered perovskite compound Ca$_2$RuO$_4$, which exhibits a paramagnetic metal-insulator transition associated with a structural transition. We showed that this recently developed {\it ab-initio} approach predicts the physically correct solutions for the two crystal structures, and electronic properties in remarkably good agreement with the available experimental data. 
A similar level of agreement has been obtained in previous LDA+DMFT studies,\cite{Gorelov2010,Sutter2017,Ricco2018,Hao2020} but these calculations involve numerous freely-adjustable parameters (Hubbard $U$, Hund coupling $J$, possible double-counting corrections, ...) which allow to fine-tune the simulation results to the experimental reference data. In the present scheme, the user merely selects the strongly correlated subspace (here, the $t_{2g}$ manifold of Ru), while all the remaining calculations are performed in a fully {\it ab-initio} manner. The level of accuracy demonstrated in this work is therefore a remarkable outcome and a very encouraging result for the further application and development of the $GW$+EDMFT framework.  

On the technical side, we have explained how materials with distorted structures and several correlated atoms within a unit cell can be efficiently treated by transforming the EDMFT impurity problems to the crystal-field basis. We also demonstrated how the frequency dependence of the interaction parameters yields significantly stronger correlations than in the static approximation, while nonlocal screening processes (introduced via the $GW$ polarization) substantially reduce the self-consistently computed interaction parameters. The subtle balance between these two effects, plus the band-widening effect of the $GW$ self-energy, are essential for reproducing the correlation effects in the material. If one of these elements is missing, as is the case for example in the EDMFT or DMFT treatment, physically incorrect solutions (e.~g. a large gap insulating solution for the $L$-Pbca structure) are obtained. 

The important role of Hund correlations has been revealed by the local state statistics, in agreement with the current understanding of the Ca$_2$RuO$_4$ phenomenology. In the insulating phase that $GW$+EDMFT predicts for the $S$-Pbca crystal structure, we obtained momentum resolved and local spectra in good agreement with available photoemission data. In the $L$-Pbca setup, where the solution is metallic, the self-consistently determined bosonic fields provided estimates for the energies associated with plasmonic satellite structures, while the analysis of the photo-emission spectra and interaction tensors allowed us to identify a lower Hubbard band feature. Finally, the Fermi surface of the metallic compound has been mapped out and agrees, up to small details, with the available photo-emission data. To illustrate the momentum-dependence of the quasi-particle renormalization introduced by the $GW$ self-energy, a map of $Z_{\bf k}$ in the Brillouin zone has been provided.

The results presented in this work demonstrate the reliability of $GW$+EDMFT for the {\it ab-initio} description of complex transition metal compounds, which are located in the computationally challenging regime of intermediate correlation strength, where for example small changes in the crystal structure can trigger a metal-insulator transition. The method provides important insights into the interplay between correlations and screening, and allows to quantify the effective interaction strength in solids, which is one of the main challenges in the theoretical study of correlated materials.
%
%
%
%
\begin{acknowledgments}
We thank Antoine Georges for proposing Ca$_2$RuO$_4$ as a benchmark material for $GW$+EDMFT, and for insightful discussions. We also thank Fredrik Nilsson and Ferdi Aryasetiawan for helpful discussions.  F.P., V.C. and P.W. acknowledge support from the Swiss National Science Foundation 
through NCCR MARVEL and SNSF Grant No~200021\_196966, and from the European Research Council through ERC Consolidator  Grant 724103. The calculations were performed on the Beo04/Beo05 clusters at the University of Fribourg. 
\end{acknowledgments}
%
%
%
%
\appendix
\section{Tensor Transformation}
The most general two-particle scattering term is given by the rank-4 tensor
\begin{equation}
\mathcal{H}=\frac{1}{2}\sum_{\sigma_{1}\sigma_{2}\sigma_{3}\sigma_{4}}\sum_{abcd}U_{acdb}\hat{c}_{a\sigma_{1}}^{\dagger}\hat{c}_{c\sigma_{2}}^{\dagger}\hat{c}_{d\sigma_{3}}\hat{c}_{b\sigma_{4}},
\end{equation}
which, by imposing spin conservation $\sigma_{1}\neq\sigma_{2}$,  $\sigma_{3}\neq\sigma_{4}$ and relabeling the orbital indexes, becomes equivalent to the effective local interaction used in $GW$+EDMFT,
\begin{equation}
\mathcal{U}\left(\omega\right)=\sum_{abcd}\mathcal{\mathcal{U}}_{abcd}(\omega)\hat{c}_{a\uparrow}^{\dagger}\hat{c}_{b\uparrow}\hat{c}_{c\downarrow}^{\dagger}\hat{c}_{d\downarrow}.
\end{equation}
Considering the specific case of a three-orbital system, the product basis representation of $\mathcal{U}$ at a given frequency can be decomposed in terms of Kronecker products between $3\times3$ density-matrix-like operators $\hat{\rho}_{\left(ab\right)}$: 
\begin{align}
 & \mathcal{\mathcal{U}}_{abcd}\left(\omega\right)\left[\hat{\rho}_{\left(ab\right)}\otimes\hat{\rho}_{\left(cd\right)}\right]= \notag \\
 & \mathcal{\mathcal{U}}_{abcd}\left(\omega\right)\left(\begin{array}{ccc}
\rho_{11}\left[\begin{array}{ccc}
\rho_{11} & \rho_{12} & \rho_{13}\\
\rho_{21} & \rho_{22} & \rho_{23}\\
\rho_{31} & \rho_{32} & \rho_{33}
\end{array}\right] & \rho_{12}\left[...\right] & \rho_{13}\left[...\right]\\
\rho_{21}\left[\begin{array}{ccc}
\rho_{11} & \rho_{12} & \rho_{13}\\
\rho_{21} & \rho_{22} & \rho_{23}\\
\rho_{31} & \rho_{32} & \rho_{33}
\end{array}\right] & \rho_{22}\left[...\right] & \rho_{23}\left[...\right]\\
\\
\rho_{31}\left[...\right] & \rho_{32}\left[...\right] & \rho_{33}\left[...\right] \label{eq:decomp}
\end{array}\right).
\end{align}
In the original (orbital) basis the $\hat{\rho}_{\left(ab\right)}$ operators are simply given by the sum of $3\times3$ single-entry matrices $\delta_{\left[ab\right]}$,
\begin{equation}
\hat{\rho}_{\left(ab\right)}=\left[\begin{array}{ccc}
1 & 0 & 0\\
0 & 0 & 0\\
0 & 0 & 0
\end{array}\right]+\left[\begin{array}{ccc}
0 & 1 & 0\\
0 & 0 & 0\\
0 & 0 & 0
\end{array}\right]+\ldots=\sum_{ab}\delta_{\left[ab\right]},
\end{equation}
and the nonzero matrix elements of Eq.~\eqref{eq:decomp} can be selected by grouping the indices as $\mathcal{\mathcal{U}}_{(ab)(cd)}(\omega)$. In order to transform the tensor to a different orbital basis defined by the $3\times3$ rotation $\Theta\in\mathrm{SO}\left(3\right)$, one separately rotates all the nine contributions to $\hat{\rho}_{\left(ab\right)}$ so that
\begin{equation}
\tilde{\rho}_{\left(ab\right)}=\sum_{ab}\Theta^{T}\delta_{\left[ab\right]}\Theta .
\end{equation}
Then the Kronecker product is performed using the rotated $\tilde{\rho}_{\left(ab\right)}$. This procedure yields the matrix representation of the interaction tensor in the product basis, which corresponds to the single-particle basis selected by $\Theta$. We checked that for the particular case of a rotationally invariant Kanamori interaction our implementation leaves the representation unchanged for {\it any} $\Theta\in\mathrm{SO}\left(3\right)$.
\bibliography{paper}
\end{document}